\documentclass[11pt,reprint,pre]{revtex4-1}
\usepackage[utf8]{inputenc}
\usepackage{amsmath}
\usepackage[utf8]{inputenc} 

\usepackage[english]{babel}
\usepackage{tikz}

\usepackage[letterpaper,top=2cm,bottom=2cm,left=2cm,right=2cm,marginparwidth=1.75cm]{geometry}

\usepackage{amsmath}
\usepackage{graphicx}
\usepackage[colorlinks=true, allcolors=blue]{hyperref}

\renewcommand{\k}{\rho}

\begin{document}
\author{Laurent Talon}
\email{laurent.talon@universite-paris-saclay.fr}
\affiliation{
Universit\'e  Paris-Saclay,  CNRS,  FAST,  91405,  Orsay, France
}
\author{Andreas Andersen Hennig}
\affiliation{
Universit\'e  Paris-Saclay,  CNRS,  FAST,  91405,  Orsay, France
}
\affiliation{
PoreLab, Department of Physics,\\ Norwegian University of
Science and Technology, N-7491, Trondheim, Norway
}

\author{Alex Hansen}
\affiliation{
PoreLab, Department of Physics,\\ Norwegian University of
Science and Technology, N-7491, Trondheim, Norway
}
\author{Alberto Rosso}
\affiliation{
Universit\'e  Paris-Saclay,  CNRS,  LPTMS,  91405,  Orsay, France
}

\begin{abstract}
The flow of yield stress fluids in porous media presents interesting complexity due to the interplay between the non-linear rheology and the heterogeneity of the medium. A remarkable consequence is that the number of flow paths increases with the applied pressure difference and is responsible for a non-linear Darcy law.  Previous studies have focused on the protocol where the  pressure difference is imposed. Here we consider instead the case of imposed flow rate, $Q$. In contrast to Newtonian fluids, the two types of boundary conditions have an important influence on the flow field.  Using a two-dimensional pore network model we observe a boundary layer of merging flow paths of size $\ell(Q) \sim Q^{-\mu/\delta}$ where $\mu = 0.42 \pm 0.02$ and $\delta \simeq 0.63 \pm 0.05$.  Beyond this layer the density of the flow paths is homogeneous and grows as $Q^\mu$. Using a mapping to the directed polymer model we identify $\delta$ with the roughness exponent of the polymer. We also characterize the statistics of non-flowing surfaces in terms of avalanches pulled at one end.
\end{abstract}

\title{Influence of the imposed flow rate boundary condition on the flow of Bingham fluid in porous media}

\maketitle

\section{Introduction}

The flow of non-Newtonian fluids is widely used in many geological and industrial applications. In particular,
Yield Stress Fluids (YSF), fluids that require a minimum amount of stress to flow, have significant implications for several industrial processes. They include slurries, polymers, oil, and foam suspensions, all of which may have a yielding stress \cite{coussot05}. One prominent application is in the extraction of heavy oil, a subject that has been extensively studied since the 1960s \citep{entov67}. Yield stress fluids are also used  to obstruct preferential paths, which can be useful in  enhanced oil recovery but also in limiting the spreading of contaminants in the ground.

Given the diverse applications, the flow of yield stress fluids has been the focus of numerous investigations \cite{entov67,pascal81,alfariss85,chase05,chen05,sochi08,chevalier13, talon13b, castro14, chevalier14, bleyer14, nash16, shahsavari16}.
A primary objective is to generalize the Darcy law, that relate the total flow rate to the applied pressure drop, for yield stress fluid.

From a practical point of view, there are several methods for driving flow in porous media.  When the flow is through a core sample driven by pumps, the flow rate is the natural control variable. However, when the flow is driven by a height difference between fluid reservoir and inlet, the pressure difference is the control parameter \cite{moura2020intermittent}. 
For Newtonian flow, it is usually assumed that the classical Darcy law is independent of the prescribed boundary condition (BC) at the inlet and outlet (imposing uniform pressure or velocity). The reason behind this is that changing the boundary condition only affects the flow in a very short distance, i.e., over a few pores.
It follows that if the total volume of interest is large enough, the influence of this boundary layer is negligible. The objective of this paper is to investigate the influence of the boundary condition for yield stress fluids.

We are interested here in the Bingham rheology for which the  stress $\sigma$ is related to the  shear rate $\dot \gamma$ by the relationship:
\begin{equation} 
\sigma = \sigma_y +  \eta \dot \gamma,
\end{equation}
where $\sigma_y$ is the yield stress, below which there is no flow ($\dot \gamma =0$).

By prescribing the pressure difference $\Delta P$, previous studies \cite{roux87,chevalier15a,bauer19,liu19} have demonstrated that the total flow rate exhibits three successive power-law regimes above a critical pressure $\Delta P_c$:
$Q \propto (\Delta P - \Delta P_c)^{\beta}$, where $\beta=1$ when the pressure is either close and very far from $\Delta P_c$. In the intermediate regime, the exponent $\beta$ is close to two \cite{liu19}. The physical origin of this quadratic exponent lies in a progressively increasing number of channels in which there is flow as the pressure is increased (see also \cite{waisbord19}).
Another important point for later is the distribution of lengths of the new flow path, where longer channels are more likely to flow than shorter ones \cite{liu19,schimmenti23}.
It is also worth mentioning the NMR experimental work of Chevalier \emph{et al.} \cite{chevalier14} where no fluid at rest was observed even at low flow rates. The authors then suggest that the breakdown of the non-Newtonian character may be due to the boundary condition.

Indeed, all these computational and theoretical studies were conducted with prescribed pressure at the boundaries. The aim of this research is thus to study the flow of yield stress fluids in porous media while imposing a prescribed flow rate. More specifically, a mixed boundary condition will be used: a uniform velocity is imposed at the inlet, while a uniform pressure is applied at the outlet. This corresponds to the experimental situation where the flow into the core sample is controlled by the rate of the pumps, whereas the other end of the core sample is open and hence kept at atmospheric pressure.  

We describe in Section \ref{sec2} the pore network model we use.  Section \ref{sec3} presents the numerical results from this model.  We map the problem onto the directed polymer problem in Section \ref{sec4} and in Section \ref{sec5} we identify the critical exponents observed in the flow problem with the corresponding exponents in the directed polymer problem. Section \ref{sec6} contains our summary and discussion of our results.   

\section{Numerical method}
\label{sec2}

\begin{figure}
\begin{tikzpicture}[scale=1]
    \node at (0,0) {\Large $\bullet$};
    \node at (0,1) {\Large $\bullet$};
    \node at (0,2) {\Large $\bullet$};
    \node at (0,3) {\Large $\bullet$};
    \node at (0,4) {\Large $\bullet$};
    \node at (1,0.5) {\Large $\bullet$};
    \node at (1,1.5) {\Large $\bullet$};
    \node at (1,2.5) {\Large $\bullet$};
    \node at (1,3.5) {\Large $\bullet$};
    \node at (1,4.5) {\Large$\bullet$};
    \node at (2,0) {\Large $\bullet$};
    \node at (2,1) {\Large $\bullet$};
    \node at (2,2) {\Large $\bullet$};
    \node at (2,3) {\Large $\bullet$};
    \node at (2,4) {\Large $\bullet$};
    \node at (3,0.5) {\Large $\bullet$};
    \node at (3,1.5) {\Large $\bullet$};
    \node at (3,2.5) {\Large $\bullet$};
    \node at (3,3.5) {\Large $\bullet$};
    \node at (3,4.5) {\Large$\bullet$};
    \node at (4,0) {\Large $\bullet$};
    \node at (4,1) {\Large $\bullet$};
    \node at (4,2) {\Large $\bullet$};
    \node at (4,3) {\Large $\bullet$};
    \node at (4,4) {\Large $\bullet$};
    \draw[line width=2] (0,0) -- (1,0.5);
    \draw[line width=3] (2,0) -- (3,0.5);
    \draw[line width=1.5] (0,1) -- (1,1.5);
    \draw[line width=2.5] (2,1) -- (3,1.5);
    \draw[line width=3] (0,2) -- (1,2.5);
    \draw[line width=2] (2,2) -- (3,2.5);
    \draw[line width=2.5] (0,3) -- (1,3.5);
    \draw[line width=3] (2,3) -- (3,3.5);
    \draw[line width=2.5] (0,4) -- (1,4.5);
    \draw[line width=2] (2,4) -- (3,4.5);

    \draw[line width=2.5] (1,0.5) -- (2,0.);
    \draw[line width=1] (3,0.5) -- (4,0.);
    \draw[line width=3] (1,1.5) -- (2,1.);
    \draw[line width=2] (3,1.5) -- (4,1.);
    \draw[line width=1] (1,2.5) -- (2,2.);
    \draw[line width=3] (3,2.5) -- (4,2.);
    \draw[line width=2] (1,3.5) -- (2,3.);
    \draw[line width=3] (3,3.5) -- (4,3.);
    \draw[line width=1] (1,4.5) -- (2,4.);
    \draw[line width=1] (3,4.5) -- (4,4.);
    
    \draw[line width=1.5] (0,1) -- (1,0.5);
    \draw[line width=2.5] (2,1) -- (3,0.5);
    \draw[line width=1.5] (0,2) -- (1,1.5);
    \draw[line width=3.5] (2,2) -- (3,1.5);
    \draw[line width=1.5] (0,3) -- (1,2.5);
    \draw[line width=3.5] (2,3) -- (3,2.5);
    \draw[line width=2.5] (0,4) -- (1,3.5);
    \draw[line width=1.5] (2,4) -- (3,3.5);

    \draw[line width=3.5] (1,0.5) -- (2,1);
    \draw[line width=1] (3,0.5) -- (4,1);
    \draw[line width=2.5] (1,1.5) -- (2,2);
    \draw[line width=1.5] (3,1.5) -- (4,2);
    \draw[line width=1.5] (1,2.5) -- (2,3);
    \draw[line width=3.5] (3,2.5) -- (4,3);
    \draw[line width=1.5] (1,3.5) -- (2,4);
    \draw[line width=2.5] (3,3.5) -- (4,4);
 
    \node at (-0.8,0) {$q_{in}$};   
    \draw[->] (-0.6,0) -- (-0.2,0);
    \node at (-0.8,1) {$q_{in}$};   
    \draw[->] (-0.6,1) -- (-0.2,1);
    \node at (-0.8,2) {$q_{in}$};   
    \draw[->] (-0.6,2) -- (-0.2,2);
    \node at (-0.8,3) {$q_{in}$};   
    \draw[->] (-0.6,3) -- (-0.2,3);
    \node at (-0.8,4) {$q_{in}$};   
    \draw[->] (-0.6,4) -- (-0.2,4);

    \draw[->] (4.5,0) -- (5,0);
    \node[] at (5,0.2) {$x$};
    \node[] at (4.67,0.6) {$y$};
    \draw[->] (4.5,0) -- (4.5,0.5);
    \node[anchor=west] at (4.2,2) {$P=0$};   
\end{tikzpicture}
\caption{\label{fig:sketch} Sketch of the pore network model. The radius of each link is randomly distributed. At each node belonging to the left boundary a flow rate $q_{in}$ is imposed. A pressure $P=0$ is applied to each node belonging to the right boundary. }
\end{figure}
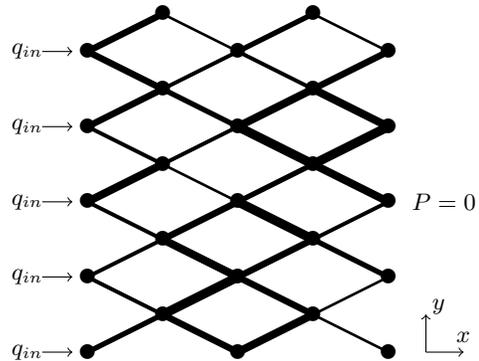

In this work we have used a Pore Network Model (PNM). As shown in Fig. \ref{fig:sketch}, the porous medium is described as an ensemble of nodes  $i=1,2, \ldots, N$ connected by links (throats). This model requires some assumptions, but it is computationally  more efficient  than the solution of the full Navier-Stokes equation inside each pore (e.g., \cite{talon13b}).
It is worth noting that such a model has been validated with direct simulations or experiments by several authors. In the context of non-Newtonian fluids, one can cite for example \cite{fraggedakis21} for the first opening path, or \cite{lopez03,sochi08} for the flow at high flow rate.

The PNM model has already been presented in detail in previous articles \cite{liu19,talon20}, but we recall the main features here: Each link is assumed to follow an approximated non-linear Poiseuille (or Buckingham-Reiner) flow. For example the flow in the link connecting pores $i$ and $j$ satisfies  
\begin{equation}
    \tilde{\delta p_{ij}} =\tilde p_i-\tilde p_j =  \k_{ij} \tilde q_{ij} + \tilde \tau_{ij} \frac{\tilde q_{ij}}{|\tilde q_{ij}|} .
    \label{eq:link_poiseuille}
\end{equation}
Here $\tilde{\delta p_{ij}}$ represents the pressure difference between nodes $i$ and $j$. The link is a tube with  radius $r_{ij}$ and  length $l_{ij}$. As a consequence, its  hydraulic resistivity  is $\k_{ij}= (8 \mu l_{ij})(\pi r_{ij}^4)$ and its pressure threshold  $\tilde{\tau_{ij}}= (2 \sigma_y l_{ij})/ r_{ij}$.

The unknown variables are the pressure at each node and the flow rate in each link. In addition to the constitutive equation eq.\ \eqref{eq:link_poiseuille} for each link, mass conservation requires $\sum_j \tilde q_{ij} =0$ at each node $i$, where the sum is over all neighbouring links $j$.
Usually two boundary conditions are considered: (i) The pressure imposed boundary condition (BC) where the pressures of all  the inlet and outlet nodes are imposed. (ii) The flow rate imposed BC, where the flow of all the inlet nodes is imposed to be equal to $\tilde q_{\text{in}}$ and the pressure of  all the outlet nodes is set to be zero. Note that in this case the pressure of the inlet nodes is in general non uniform. 

The pressure-imposed BC has been studied in \cite{talon13b,liu19,schimmenti23}. Here we study the flow rate-imposed BC.
To take into account the disorder of the system, the radius is drawn randomly from a uniform distribution with mean $r_0$ and standard deviation $\sigma_r$. The system size is $L \times W $ nodes, where $L$ is in the direction of the flow. We have set the variables to the values $r_0=0.25$, $\sigma_r=0.15$, $\sigma_y=1$, $l=1$.
In the following, we use the dimensionless flow rate and pressure variables $q = \tilde q\ {4 \mu}/{ \tau_y \pi r_0^3}$ and $P = \tilde{P}\ {r_0}/{2 \sigma_y l}$.
 
The solution of the system of nonlinear equations is a set of flow rates $\{q_{ij}\}$. The solution can be determined using a variational approach that minimize a functional \cite{talon20,talon22}. In this case it writes as
\begin{equation}
\label{eq:variational}
   \{q_{ij}\} = \arg \min_{\{q_{ij}\}\in\Omega} \sum (\frac{1}{2} \k_{ij} q_{ij}^2 + \tau_{ij} |q_{ij}|),
\end{equation}
where $\Omega$ is the ensemble of the admissible solutions satisfying conservation of mass and the prescribed flow at the inlet. Our numerical results are obtained using a modified version of this variational approach, called augmented Lagrangian method and is described in Talon and Hansen \cite{talon20}. Eq.\ \eqref{eq:variational} allows one also to obtain an important insight of the solution when $Q\rightarrow 0$ or $Q \rightarrow \infty$.

\section{Numerical results}
\label{sec3}

\subsection{Flow curve}
\begin{figure}
\includegraphics[width=0.99\hsize, trim=10 0 10 0, clip= true]{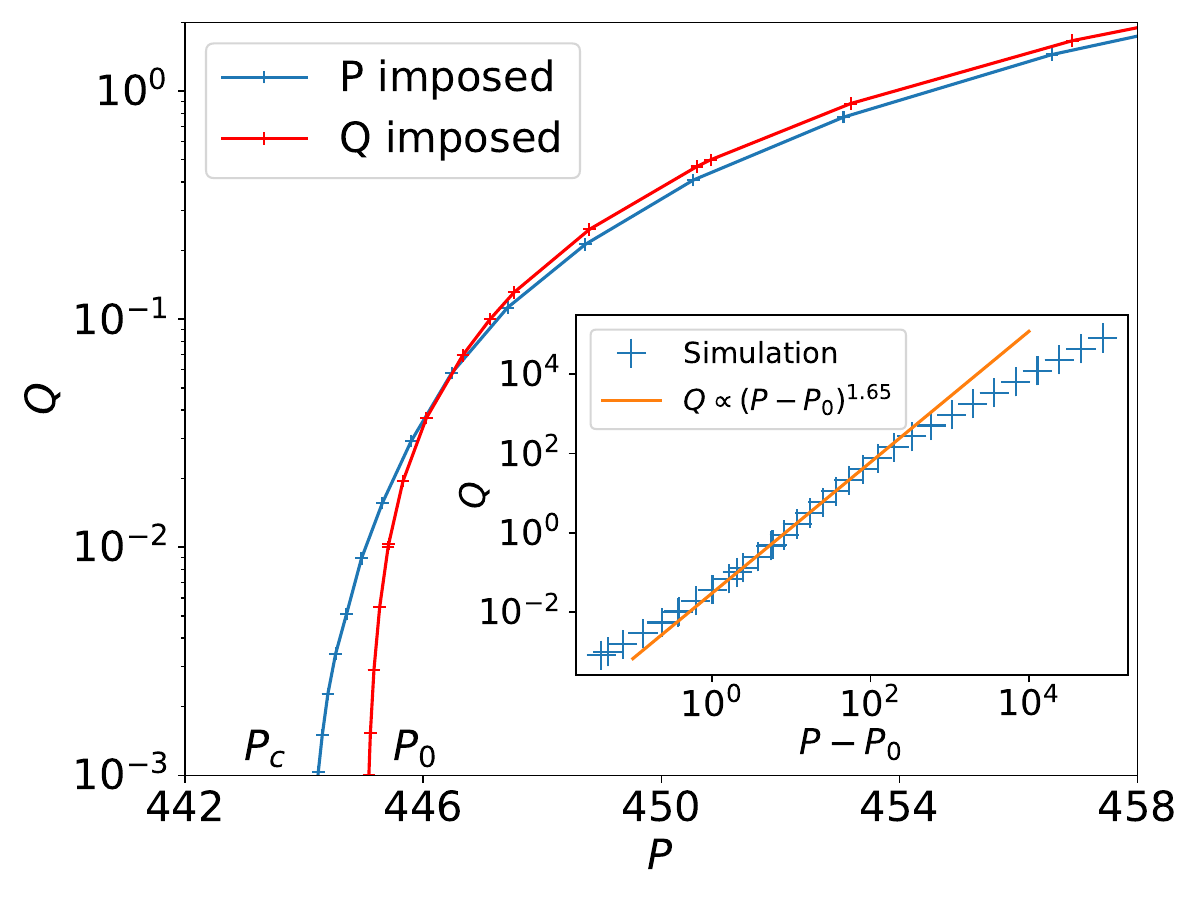}
\caption{ \label{fig:QP} Total flow rate $Q$ as a function of pressure using imposed pressure difference BC (blue) and imposed flow  rate BC (red).
The minimum pressure for the onset of flow is different for the two conditions. Inset: log-log plot of flow rate as a function of distance to onset pressure $P-P_0$ for imposed flow rate BC. The line is a power law fit. The system size is $512\times512$.
}
\end{figure}

We first compute the macroscopic flow-pressure curve as shown in  Fig.\ \ref{fig:QP}.  For both imposed pressure and imposed flow BC, we find that a minimal pressure is needed to observe a finite flow. Remarkably the minimal pressure for the imposed flow BC, $P_0$ is always larger than the one for imposed pressure difference BC, $P_c$. Moreover, as shown in the inset of Fig.\ \ref{fig:QP} and similarly to the imposed pressure difference BC \cite{liu19},  the imposed flow BC displays three flow regimes: At low and high flow rates the relationship between $Q$ and $ P - P_0$ is linear. In the intermediate regime a power law  is observed:
\begin{equation}
    Q \propto (  P - P_0)^{\beta},
\end{equation}
with $\beta \simeq 1.65$, which is smaller than the exponent measured for pressure imposed flow BC, which is close to $2$  \cite{chevalier15a,liu19}.

\subsection{Flow channels}
\begin{figure*}

\begin{tikzpicture}
\node[anchor=south west] at (0,0) {\includegraphics[width=0.32\hsize]{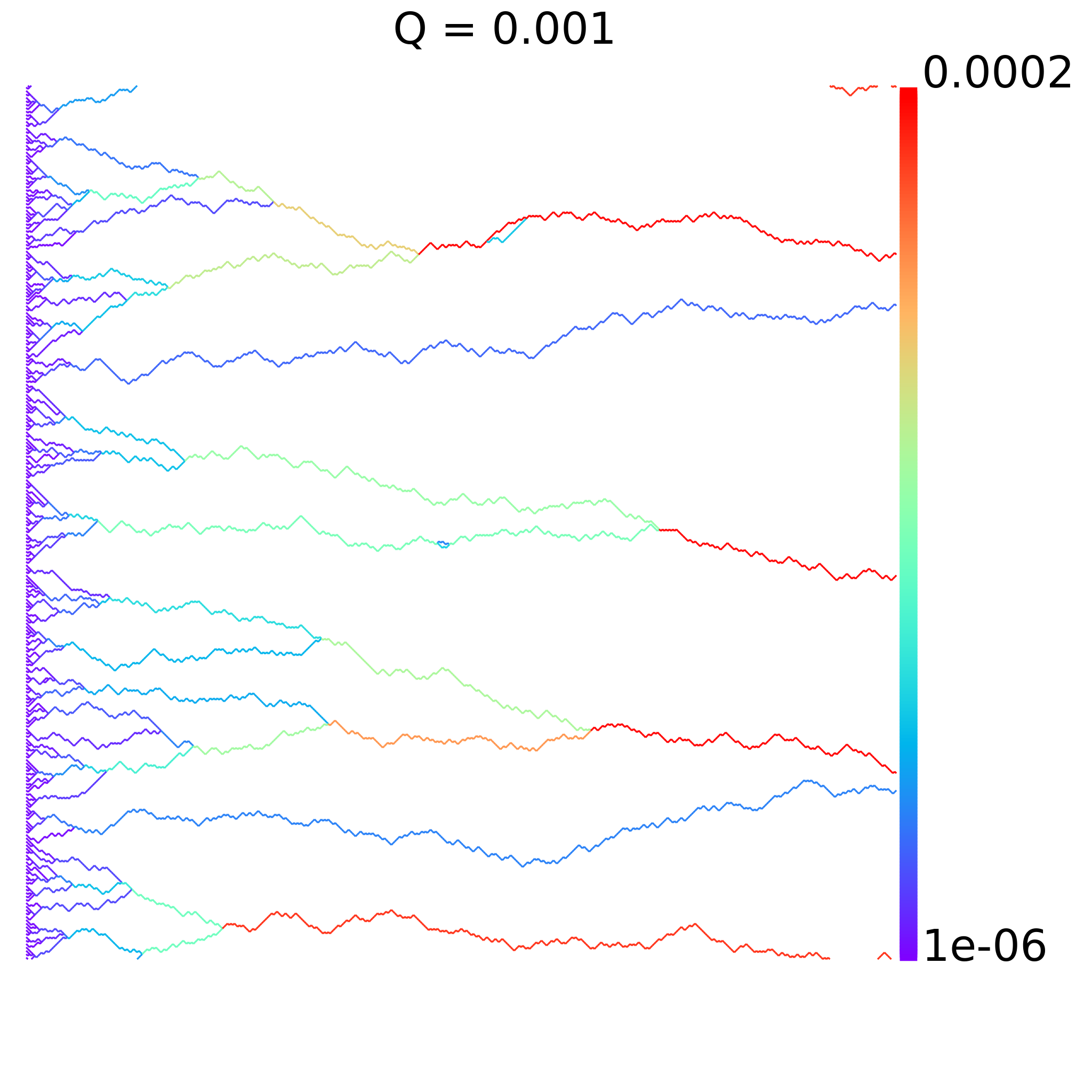}};
\node[anchor=south west] at (5.7,0) {\includegraphics[width=0.32\hsize]{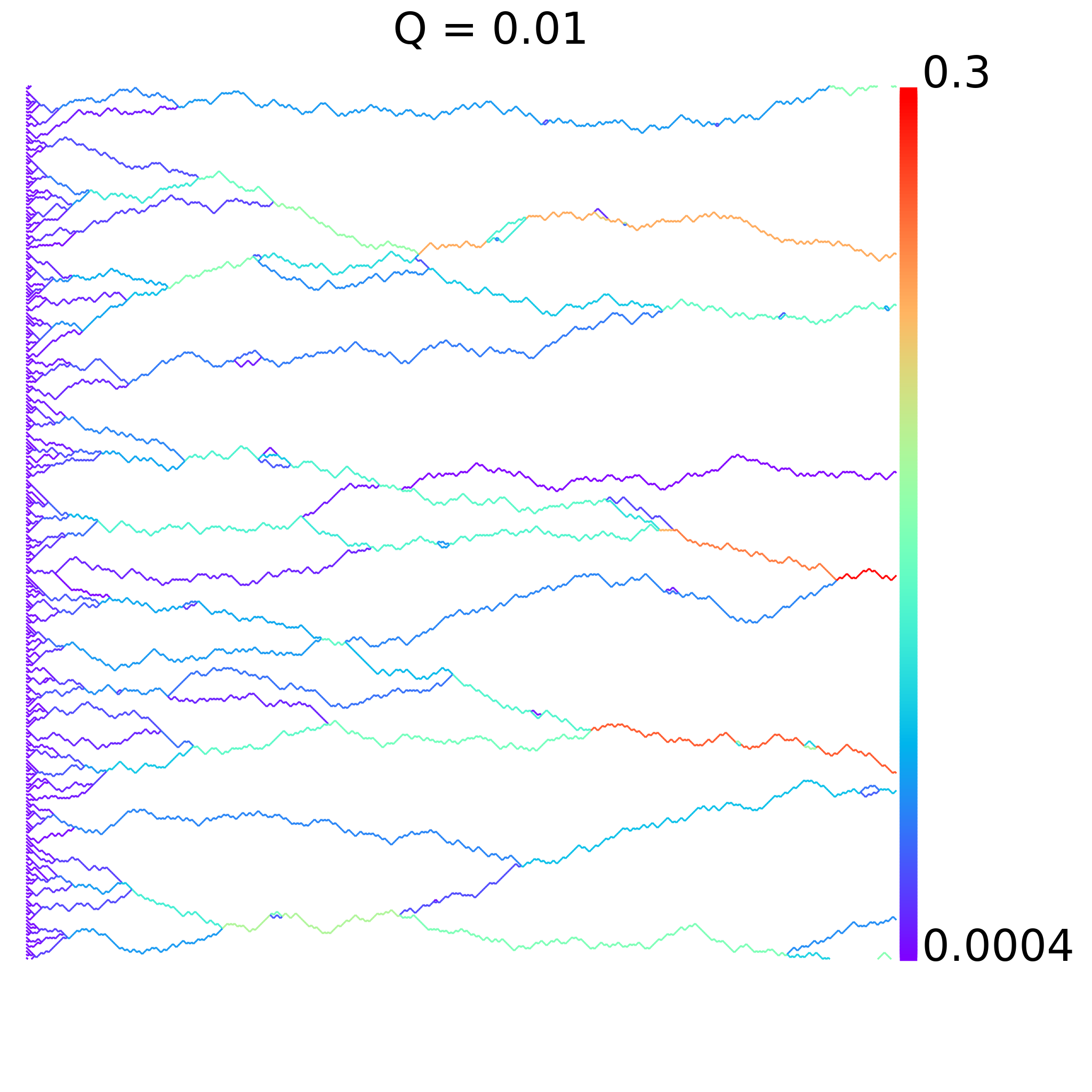}};
\node[anchor=south west] at (11.5,0) {\includegraphics[width=0.32\hsize]{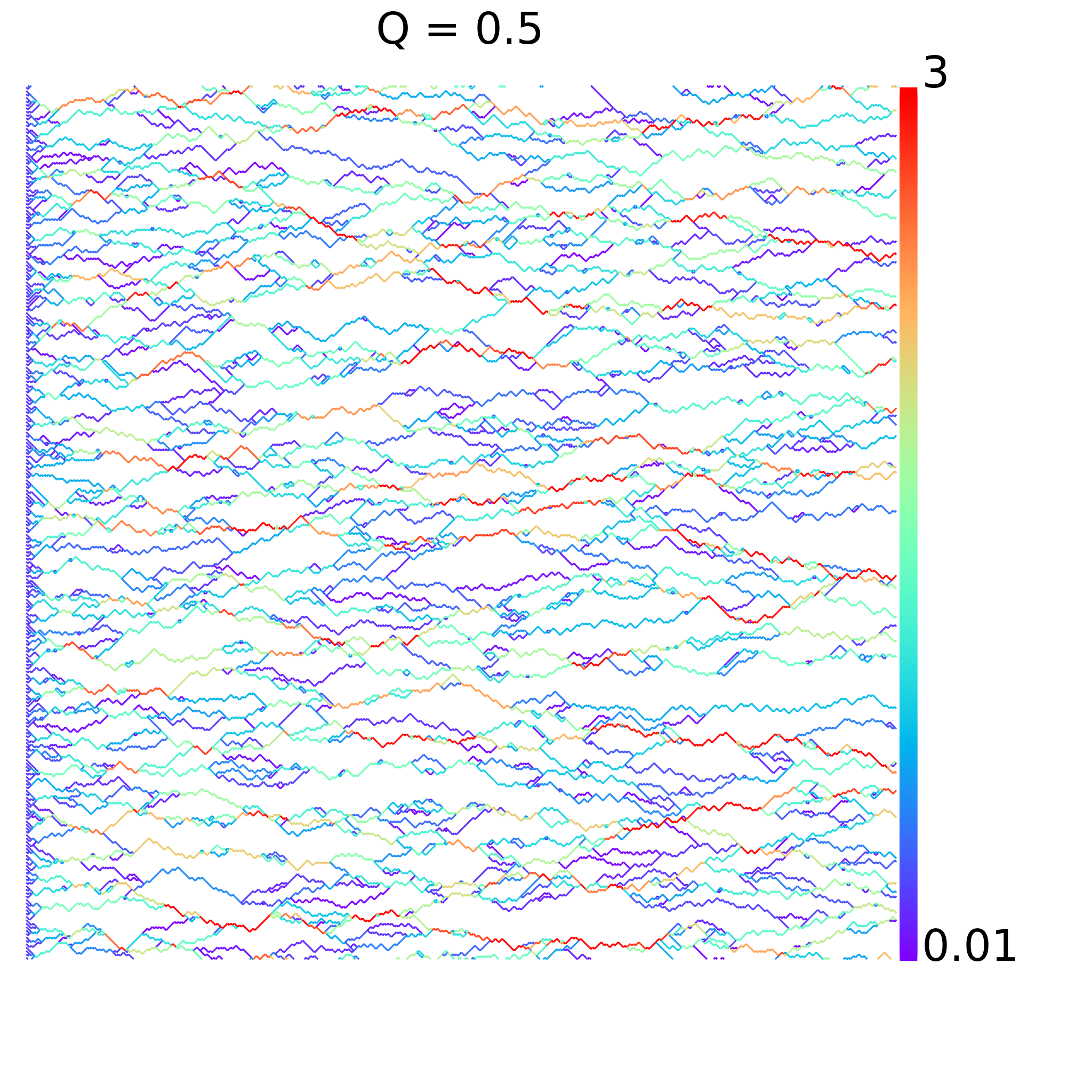}};
\draw[->,line width=2] (7,+0.5)--(9.5,+0.5);
\node[] at (8.25,0.2) {$Q$};
\end{tikzpicture}
\caption{\label{fig:flow_paths} Flow field from PNM simulations for different imposed flow rate and for a system size of $L \times W = 512\times512$.
A flow rate $Q$ is imposed at the left edge, while a pressure $P=0$ is imposed at the right edge. 
}
\end{figure*}

The origin of the flow regimes for imposed flow rate is similar to the pressure difference imposed BC and are due to the increase of paths where fluid is flowing (flowing paths). In Fig.\ \ref{fig:flow_paths}, we depict the flow behavior corresponding to different imposed total flow rate.

When the flow rate is extremely low (left panel), due to the flow imposed BC, we observe a flow path starting from each inlet node.
This is in contrast with the  pressure imposed BC, where only a single  flow path is observed. As shown in Fig.\ \ref{fig:flow_paths},
the  paths eventually merge with others as they progress through the system. The key feature  of the low flow rates is that these paths merge but never split. If the system is long enough, namely if $L\gg W$, the flow paths  eventually converge into a single channel, as for the imposed pressure difference BC.
 
As the total imposed flow rate increases (see middle panel of Fig.\ \ref{fig:flow_paths}, we  observe that the flow paths begin to split, leading to an increase in the number of flow channels downstream, which is at the origin of the power regime (similarly to the pressure imposed BC).
For very large flow rate, all the paths are open and the Newtonian limit is recovered. 

\begin{figure*}
\includegraphics[width=0.45\hsize]{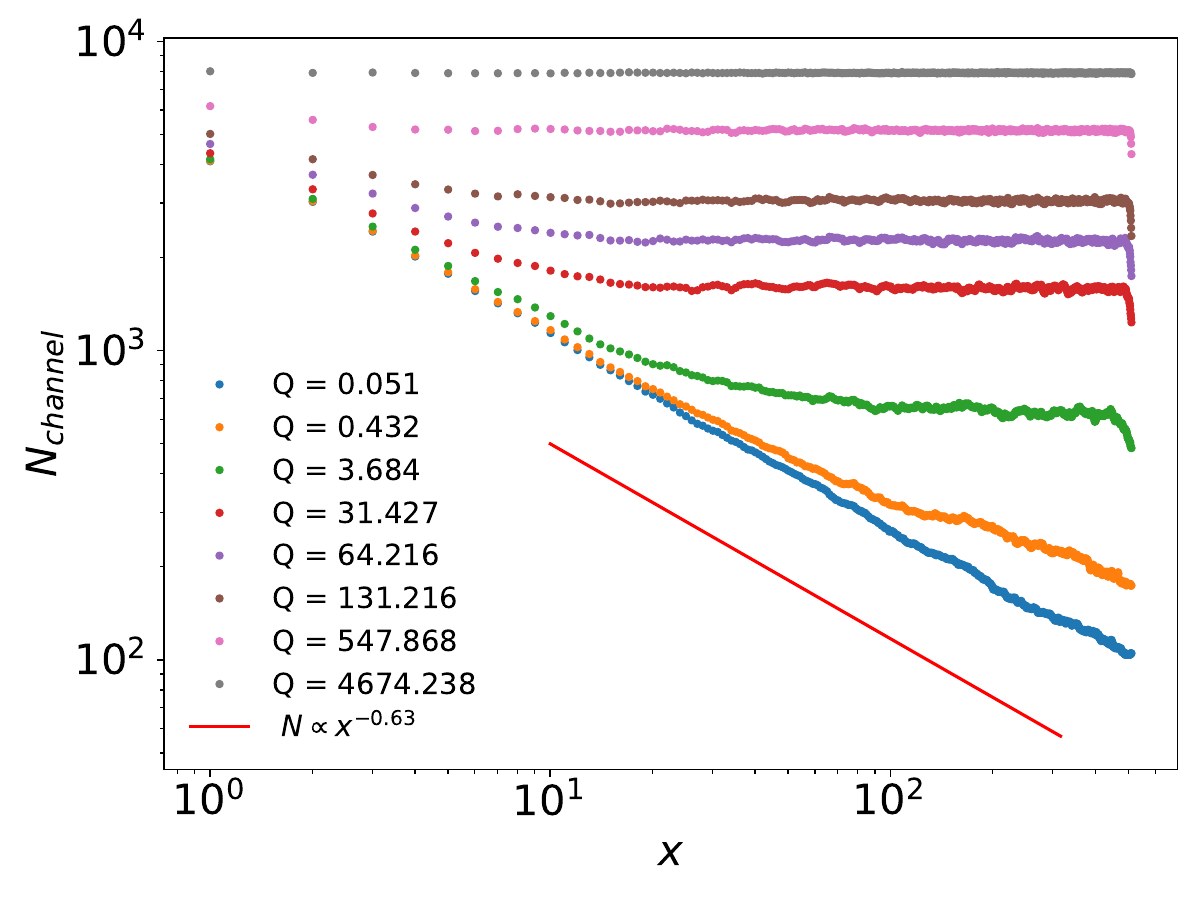}
\includegraphics[trim=0 5 0 -5,width=0.45\hsize]{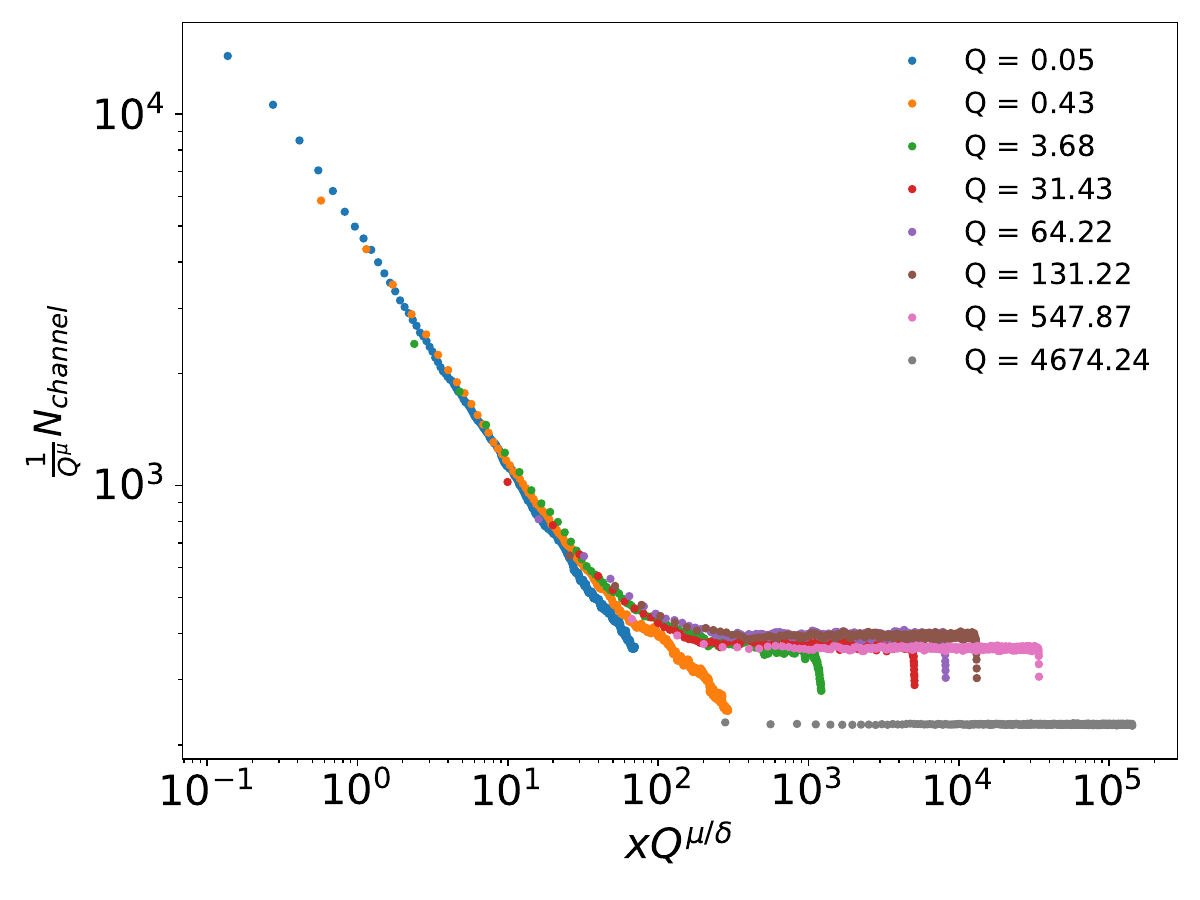}
\caption{ \label{fig:nchannel_depth} Left: Number of flowing paths as function of the depth $x$ in the flow direction for a system  of size $L \times W= 512 \times 4096$. Right: the Family-Vicsek scaling proposed in Eq.\ (\ref{eq:collapse}). We observe good collapse for $\mu = 0.42 \pm 0.02$ and $\delta \simeq 0.63 \pm 0.05$ with two exceptions: (i) close to the outlet we note a small decrease due to the imposed zero pressure condition at the boundary. (ii) For extremely large $Q$, the observed plateau is twice larger ($\approx W$) than the one predicted by the collapse  ($\approx W/2$). This is due to the number of links flowing at the inlet double at very high flow rates.
}
\end{figure*}

To quantify these trends, we plot in Fig.\ \ref{fig:nchannel_depth} (Left) the number of flowing channels $N_{\text{channel}}$ as function of the distance, $x$,  from the inlet. At low $Q$ the number of channels decreases with distance as a power law $x^{-\delta}$, with $\delta \simeq 0.63 \pm 0.05$.
\begin{equation}
    N_{\text{channel}}(x \ge 1, Q\to 0) \simeq \frac{W}{2} x^{-\delta}.
    \label{eq:Nchannel_x}
\end{equation}
As the flow rate is increased, the power law decay holds within a boundary layer of length $\ell_{\text{BL}}(Q)$. Above this length,  $N_{\text{channel}}$  reaches a plateau that grows as $Q^{\mu}$. Note that the existence of a plateau is a necessary condition to enable a homogeneous description of the flow. Also, a Family-Vicsek scaling \cite{family85} holds for  $N_{\text{channel}}(x, Q)$ (see figure \ref{fig:nchannel_depth})
\begin{equation}
    \label{eq:collapse}
    N_{\text{channel}}(x, Q) = Q^{\mu} f(x Q^{\mu/\delta})\;,
\end{equation}    
with  $\mu = 0.42 \pm 0.02$ and $\delta \simeq 0.63 \pm 0.05$. Function $f$ scales as $f(z) \sim z^{-\delta}$  for small $z$ and reaches a constant value at large $z$. 
It follows that the the boundary layer decreases with the flow rate as $\ell_{\text{BL}}(Q) \sim Q^{- \mu/\delta}$.
In the following, we will establish a mapping between the problem of the onset of the flow (very low rate regime) and the model of the directed polymer in random media. This mapping will allow us to predict the exponent $\delta$ and explain the different minimal pressure observed by changing the BC.

\section{Onset of the Flow: mapping to the directed polymer problem}
\label{sec4}

Using the flow equations we can write an equation for the energy dissipation :
\begin{equation}
  Q \Delta P =  \sum q_{ij} \delta p_{ij} =   \sum  \left( \k_{ij} q_{ij}^2 + \tau_{ij} |q_{ij}|\right)
  \end{equation}
This equation can be recast in the following form:
\begin{equation}
Q = \kappa^*(Q) \left[ \Delta P - P^*(Q) \right], 
\end{equation}
With 
\begin{equation}
     \kappa^*(Q) = \left( \frac{1}{Q^2} \sum \k_{ij} q_{ij}^2 \right)^{-1}
     \label{eq:kappa}
\end{equation}
and
\begin{equation}
  P^*(Q) = \frac{1}{Q} \sum \tau_{ij} |q_{ij}|.  
  \label{eq:P_star}
\end{equation}
Here $\kappa(Q)$ is the effective mobility and  $P^*(Q)$ is the apparent pressure threshold because it tends to the minimal threshold pressure when $Q \to 0$.

\begin{figure*}
\includegraphics[width=0.45\hsize]{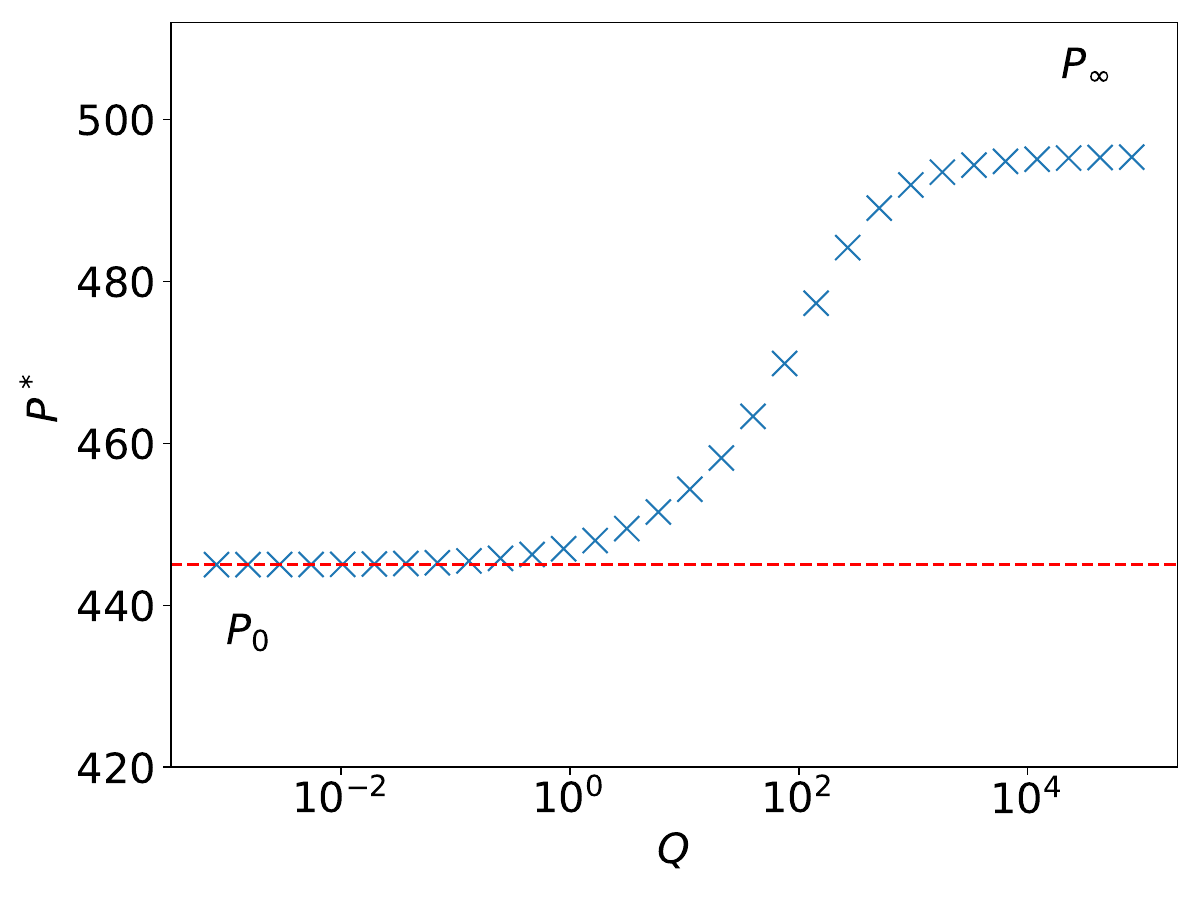}
\includegraphics[width=0.45\hsize]{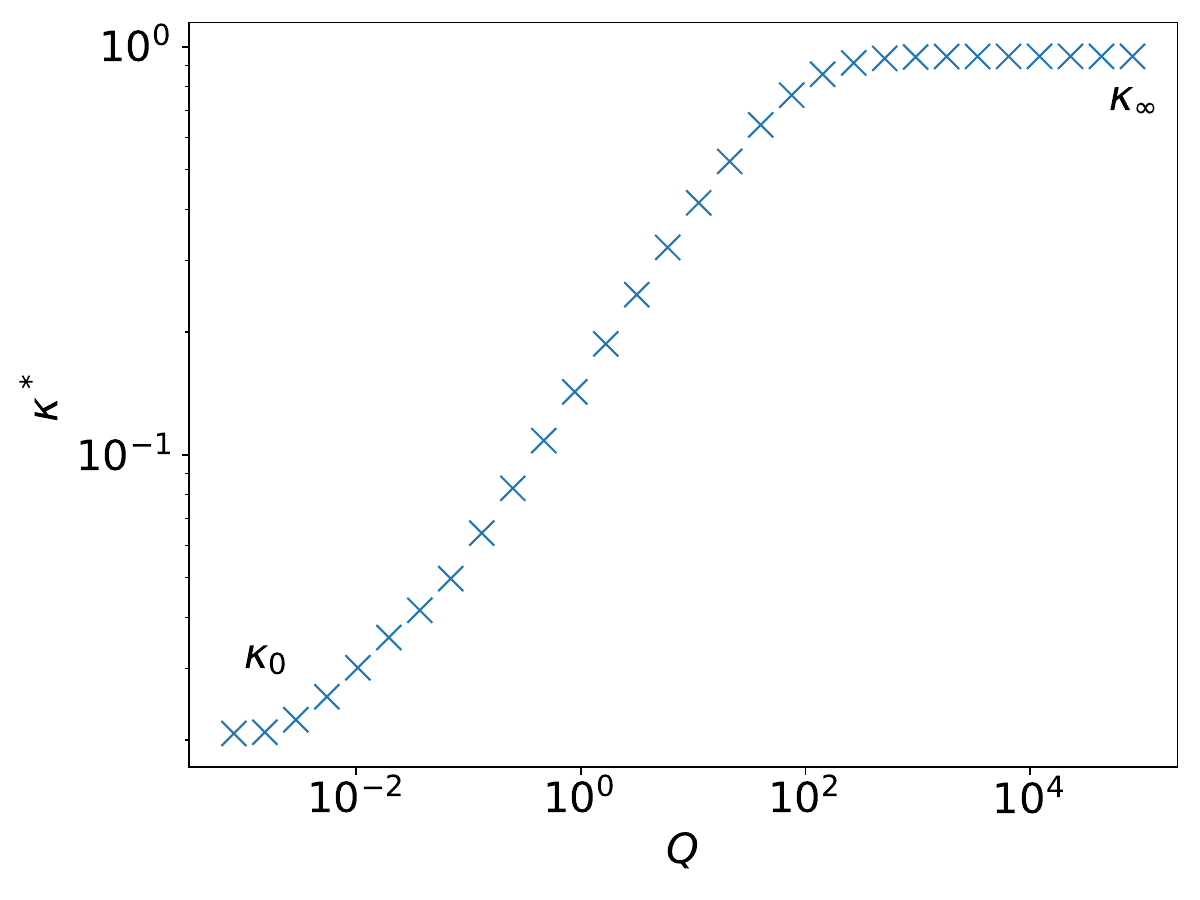}
\caption{ \label{fig:P_Q_star}(Left) Apparent pressure threshold $P^*(Q)$ as function of $Q$, dashed line is the theoretical prediction from the directed polymer problem described below. (Right) Effective mobility $\kappa^*(Q)$ as function of $Q$.}
\end{figure*}

\begin{figure*}
\begin{tikzpicture}[scale=0.5]

\draw[fill=red!20] (0,2) -- (1,3) -- (2,2) -- (3,1) -- (4,2) -- (5,1) -- (6,2) -- (7,3)-- (8,2) -- (9,1-0.1) --
(8,0-0.2) -- (7,-1-0.2) -- (6,-2-0.2) -- (5,-1-0.2)-- (4,0-0.2)-- (3,-1-0.2)-- (2,0-0.2)-- (1,1-0.2) -- (0,0-0.2);

\draw[line width=1.5pt] (0,0-0.2) -- (1,1-0.2) -- (2,0-0.2) -- (3,-1-0.2) -- (4,0-0.2) -- (5,-1-0.2) -- (6,-2-0.2) -- (7,-1-0.2)-- (8,0-0.2) -- (9,1-0.2)  -- (10,0-0.2)  -- (11,1-0.2) -- (12,0-0.2)-- (13,-1-0.2)-- (14,0-0.2)-- (15,1-0.2)-- (16,2-0.2)-- (17,1-0.2)-- (18,2-0.2) -- (19,1-0.2)-- (20,2-0.2)-- (21,3-0.2)-- (22,2-0.2)-- (23,3-0.2) -- (24,4-0.2);

\draw[line width=1.5pt] (0,2) -- (1,3) -- (2,2) -- (3,1) -- (4,2) -- (5,1) -- (6,2) -- (7,3)-- (8,2) -- (9,1)  -- (10,0)  -- (11,1) -- (12,0)-- (13,-1)-- (14,0)-- (15,1)-- (16,2)-- (17,1)-- (18,2) -- (19,1)-- (20,2)-- (21,3)-- (22,2)-- (23,3) -- (24,4);

\draw[line width=1.5pt] (0,4+0.2) -- (1,5+0.2) -- (2,6+0.2) -- (3,5+0.2) -- (4,4+0.2) -- (5,5+0.2) -- (6,6+0.2) -- (7,5+0.2)-- (8,6+0.2) -- (9,5+0.2)  -- (10,4+0.2)  -- (11,3+0.2) -- (12,4+0.2)-- (13,3+0.2)-- (14,2+0.2)-- (15,3+0.2)-- (16,4+0.2)-- (17,3+0.2)-- (18,2+0.2) -- (19,1+0.2)-- (20,2+0.2)-- (21,3+0.2)-- (22,2+0.2)-- (23,3+0.2) -- (24,4+0.2);

\node[anchor=north east] at (2,0.) {\Large $q_i$};
\node[anchor=north east] at (2,2.2) {\Large $q_i$};
\node[anchor=north east] at (2,5.3) {\Large $q_i$};

\node[anchor=north east] at (12,0) {\Large $2q_i$};
\node[anchor=north east] at (22,4.5) {\Large $3q_i$};

\node[anchor=center] at (0,0-0.2) {\Large$\bullet$};
\node[anchor=center] at (0,2) {\Large$\bullet$};
\node[anchor=center] at (0,4+0.2) {\Large$\bullet$};

\draw[<-] (-0.2,4.1) arc (95:265:1);
\draw[<-] (-0.2,1.9) arc (95:265:1);

\end{tikzpicture}
\caption{\label{fig:sketch_merging} 
Sketch of directed polymers starting from each inlet node and ending at any outlet node. Since each inlet node is separated by $1$ and each polymer is independent, the system is equivalent to the avalanche dynamics of a directed polymer pulled at the inlet end.
}
\end{figure*}
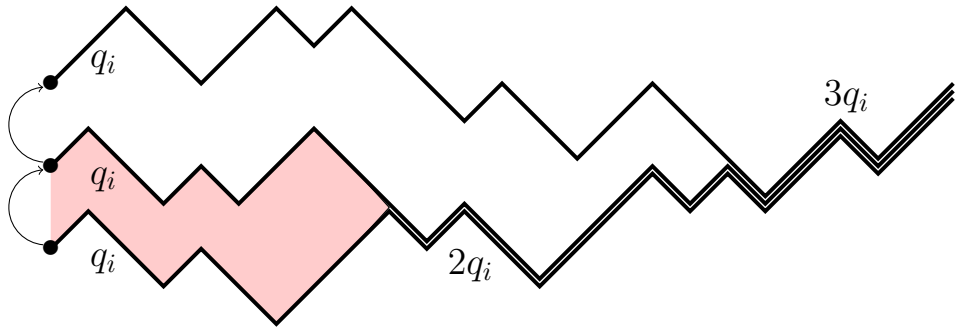

\begin{figure}
\includegraphics[width=0.9\hsize]{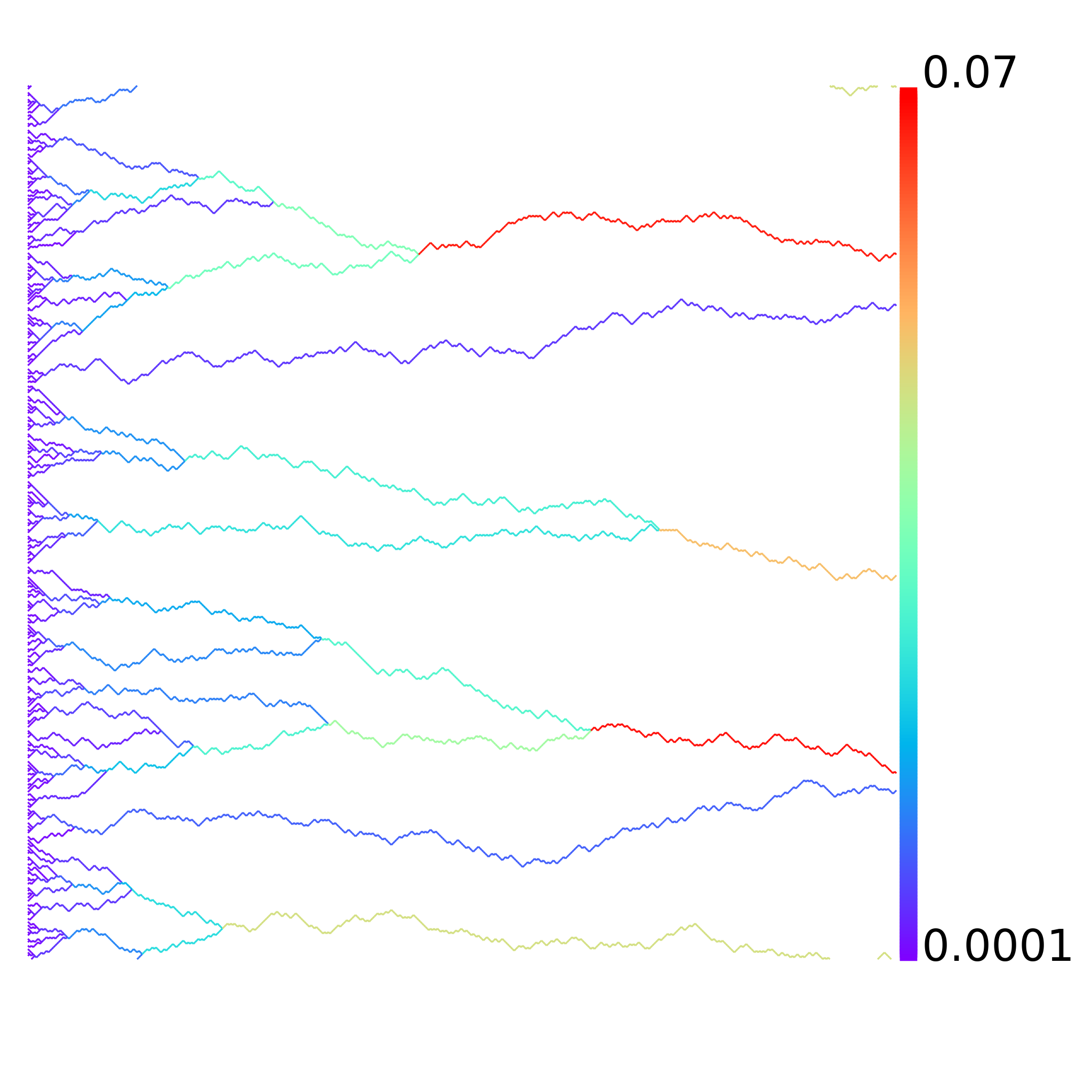}
\caption{\label{fig:groundstate} Flow field at $Q \rightarrow 0$ obtained from the directed polymer model. Large portion of non flowing surfaces are bounded by flowing channels. Their statistics is scale free $P(s) \sim S^{-\tau_S}$.}
\end{figure}

Fig.\ \ref{fig:P_Q_star} shows the evolution of $P^*(Q)$ and $\kappa^*(Q)$. Both functions increase from a plateau value, resp. $P_0$ and $\kappa_0$ at low $Q$ to a higher one, respectively $P_\infty$ and $\kappa_\infty$. This evolution is due to the increase of the number of flowing channels with $Q$.  It is possible to find an explicit expression for $P_0$ and $\kappa_\infty$ using the variational approach of Eq.\ \eqref{eq:variational}.
\begin{itemize}
\item For $Q\rightarrow \infty $, the flow field converges to the solution $\{q\} = \arg \min_{\{q\}\in\Omega} \sum (\frac{1}{2}  \k_{ij} q_{ij}^2),$ equivalent to the Newtonian flow.
Hence,  the mobility $\kappa_\infty$ corresponds to the Newtonian mobility, namely the maximal admissible mobility:
\begin{equation}
\kappa_\infty = \lim_{Q\rightarrow \infty} \max_{\{q\}\in\Omega} \left( \frac{1}{Q^2}  \sum  \k_{ij} q_{ij}^2 \right)^{-1}.
\end{equation}
\item For $Q\rightarrow 0$, the flow field converges to the solution $\{q\} \rightarrow \arg \min_{\{q\}\in\Omega_Q} \sum (\delta p_c |q|).$ 
It follows an exact expression for $P_0$:
\begin{equation}
    P_0 =  \lim_{Q\rightarrow 0}  \min_{\{q\}\in\Omega} \sum \tau_{ij} \frac{|q_{ij}|}{Q} 
    \label{eq:P0}
\end{equation}
\end{itemize}
This expression allows us to map the problem of the onset of the flow with  the directed polymer model. For flow imposed BC with identical flow rates $q_i$ for all the inlet nodes, the flow structure consists of directed flow paths that start from all inlet nodes.
Indeed, Eq.\ \eqref{eq:P0} prescribes that the flow rate of a particular link is equal to $n_{\text{paths}} q_{\text{in}}$, where $n_{\text{paths}}$ corresponds to the total number of paths passing to this link (see  Fig.\  \ref{fig:sketch_merging}). Hence, the sum in Eq.\ \eqref{eq:P0} can be rewritten as the sum of the threshold along each paths, carrying a flow rate $q_{\text{in}}$ divided by the total number of channels. Each path can be considered as a directed polymer with energy given by the sum of its thresholds. As a consequence the minimal pressure $P_0$ correspond to the mean value of the ground state energy of the $W$ directed polymers starting at each inlet node. On the other hand, for pressure imposed BC, a single channel is expected, and the minimal pressure $P_c$ corresponds to the minimal ground state energy among the $W$ polymers previously considered. It is then clear that we expect $P_c < P_0$. 
In Fig.\ \ref{fig:groundstate} we plot the flow field obtained using the algorithm of \cite{drossel95} to determine the directed polymers. The structure of the flow paths is identical to the one found at low flow rate in Fig.\  \ref{fig:flow_paths} left.
Moreover, the corresponding mean value of the ground state energies  (horizontal line in Fig.\ \ref{fig:P_Q_star}) correctly predicts the value of $P_0$ in Fig.\ \ref{fig:QP} (left).

\section{Critical exponents and  non-flowing regions}
\label{sec5}

In the previous section we identify the flowing channels at the onset of the flow with the ground states directed polymers in the random (porous) medium. This mapping allows us to use an efficient algorithm for the flowing channels, but it also allows us to identify the critical exponent $\delta$ previously introduced with the roughness exponent $\zeta=2/3$ of the directed polymer. To do this we first characterize the statistics of the non-flowing regions using a scaling relation originally introduced for the Oslo sandpile model  \cite{paczuski96a}.
This argument is applicable to avalanche dynamics of elastic interfaces in random media pulled at one end \cite{aragon16}.

In particular, from Fig.\ \ref{fig:groundstate}, we consider the non-flowing portions bounded by flowing channels. Our goal is to determine their surface, $S$, their length $\ell$, and their width $w$. 
Similar to avalanches, these portions display a scale free statistics,
\begin{equation}
    \pi_1(S) = S^{-\tau_S} g_1(S/S_{\max}),  \; \pi_2(\ell)  = \ell^{-\tau_\ell} g_2(\ell/\ell_{\max})\;.
\end{equation}
Here $\tau_S, \tau_\ell$ are the avalanche exponents with a value in the interval $(1,2)$. Both distributions have a finite size cut-off ($S_{\max} \sim L^{1+\zeta}$, with $\zeta=2/3$,  and $\ell_{\max}\sim L$), above which both $g_1(x)$ and $g_2(x)$ have a fast decay. We can determine the mean avalanche surface, $\langle S\rangle$ by  
\begin{equation}
    \langle S\rangle  \sim S_{\max}^{2-\tau_s} \sim L^{(1+\zeta)(2-\tau_S)}\;.  
\end{equation}
 To close this relation we adapt the same argument employed to characterize depinning avalanches driven at a tip \cite{aragon16} 
and originally introduced in the context of the Oslo sandpile model where grains are added on one side only  \cite{paczuski96a}. Indeed here we are determining $W$ ground state polymers. 
The first polymer is constrained to start at the inlet $(x,y)=(0,0)$, the last one at the inlet $(x,y)=(0,W)$. The total surface bounded by these two channels is $\approx W L$ and we can write
\begin{equation}
W \langle S \rangle \approx W L \Longrightarrow L^{(1+\zeta)(2-\tau_S)} \approx L\;,
\end{equation}
from which we get $\tau _S=2-1/(1+\zeta)$. The exponent $\tau_\ell$ can be derived using $S\sim \ell^{1+\zeta}$ and $S^{-\tau_S} dS \sim \ell^{-\tau_{\ell}} d \ell$. From this we get  $\tau_{\ell}=1+\zeta$. Let us finally remark that the number of channel at a given distance $N_{channel}(x)$ corresponds to the number of avalanches whose length is longer than $x$, thus
\begin{equation}
    N_{channel}(x)  \sim  W \, \text{Prob}(\ell>x) \sim  W x^{-\tau_{\ell} +1}\;.
\end{equation}
Hence, we can identify $\delta =\tau_{\ell}-1=\zeta$, which is consistent with the observation.

\begin{figure*}
    \includegraphics[width=0.32\hsize]{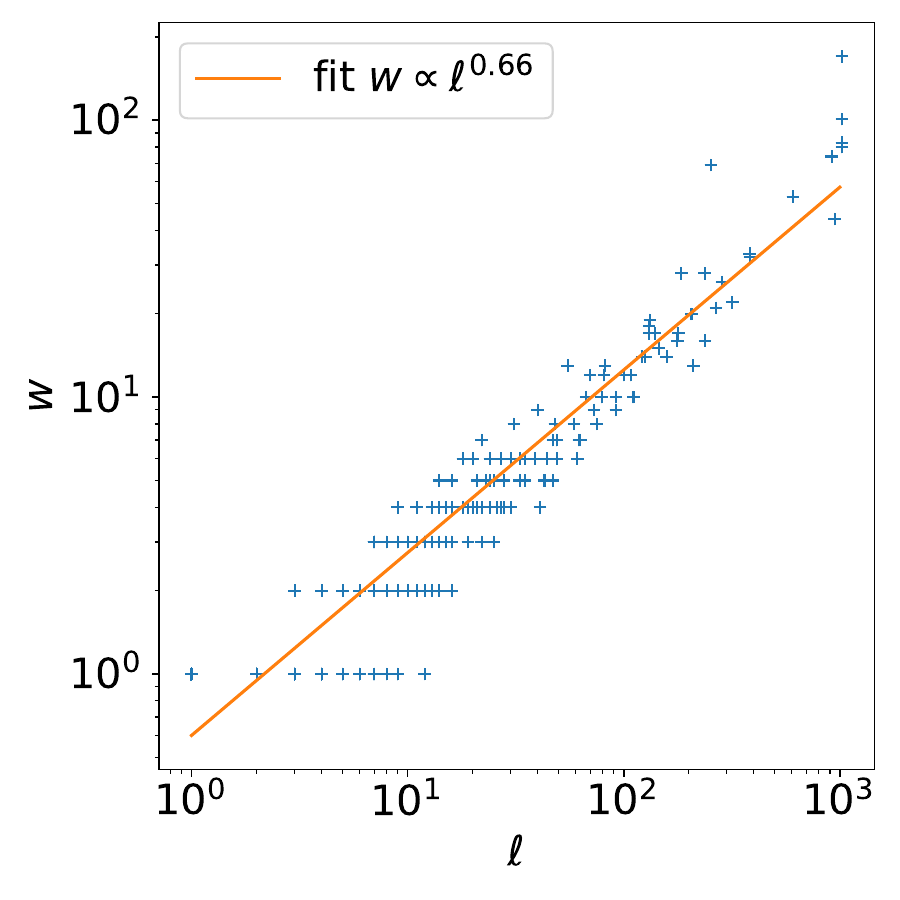}   
    \includegraphics[width=0.32\hsize]{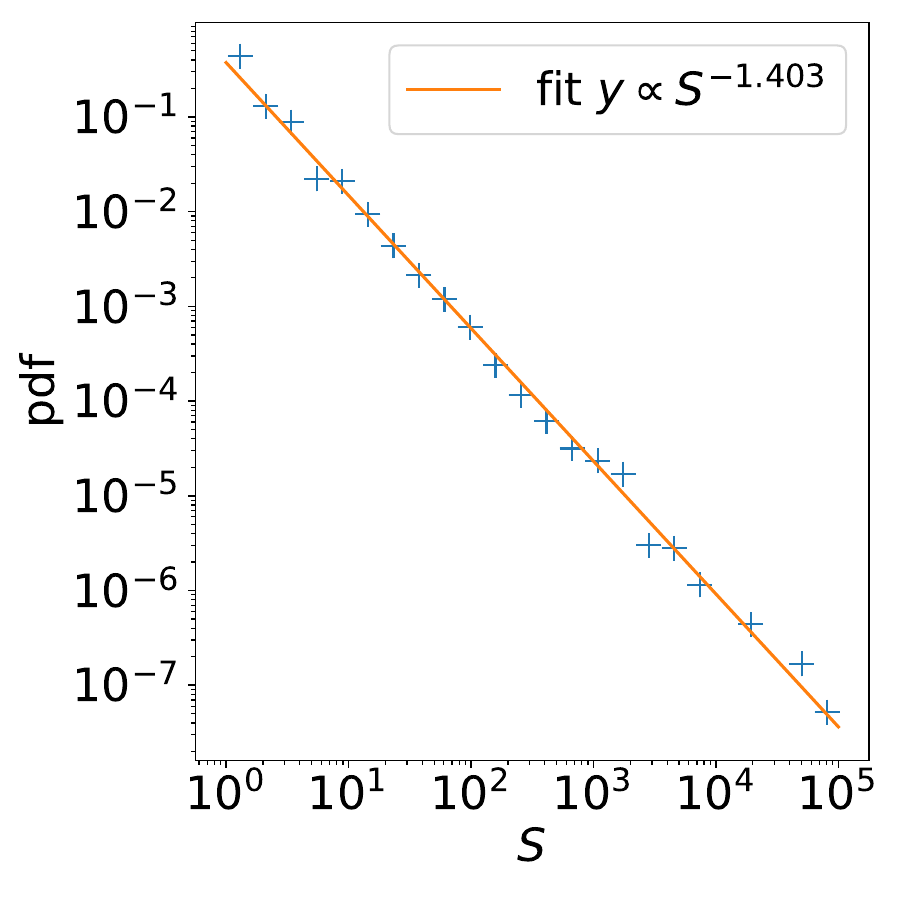}
    \includegraphics[width=0.32\hsize]{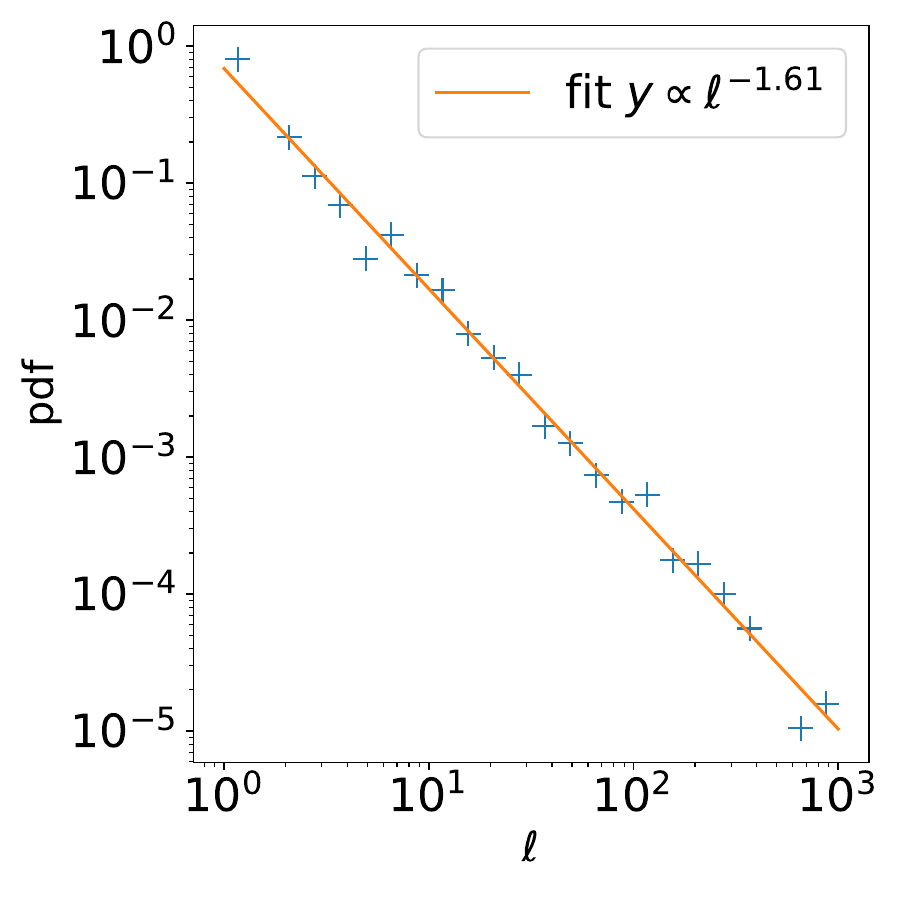}
 \caption{\label{fig:PSL} (Left) Avalanche width $w$ as function of its length $\ell$. The line represents the scaling $w\sim \ell^{\zeta}$.  Probability distribution of surface (Middle) and length (Right) of non-flowing surfaces. Here, the system size is $L \times W= 1024\times1024$.}
\end{figure*}

To summarize, the mapping with the ground state polymers allows us to determine
\begin{equation}
\tau _S=2-\frac{1}{1+\zeta}= \frac{7}{5}, \; \tau_{\ell}=1+\zeta=\frac{5}{3}, \; \delta=\zeta=2/3\;.
\end{equation}
These exponents are in good agreement with those measured in Fig.\ \ref{fig:PSL}.
It is worth noting that such a mapping cannot determine $\beta$ and $\mu$ because they depend on the channels opening at higher flow rates.

It is also worth mentioning that at very low flow rate, the power-law relationship Eq.\ \eqref{eq:Nchannel_x} indicates that the boundary layer should be infinite.
However, due to the finite size of the system, a plateau is necessarily reached once $N_{channel}=1$.
It follows a maximum length $\ell_{bc}^{max}$ for the boundary layer that depends on the domain width is
$   \ell_{bc}^{max} \propto W^{1/\delta}.$

This suggests that in order to observe the flow path beyond the boundary layer for low $Q$, one needs to consider a domain with an aspect ratio other than one ($L>W$), unlike the example in Fig.\ \ref{fig:flow_paths}.

\section{Discussion/Conclusion}
\label{sec6}

In this paper we have analysed the influence of the type of boundary condition on the flow of Bingham fluid in porous media.
In contrast to Newtonian fluids, the type of the applied boundary condition has a very strong influence on the flow, especially at low flow rates. 
One consequence is the difference in the minimal pressure for the onset of flow which is smaller when imposing the pressure imposed BC than when imposing the flow rate.
Another consequence is that, at low velocity, the pressure imposition shows a single flowing channel, whereas the flow imposition displays a merging channel structure. 
There is therefore a boundary layer that spreads over a long distance.
Its expansion increases according to a power law as the flow rate approaches zero, and increases with the system width.
It follows for instance that the boundary layer, but also the difference between the two pressure of flow onset depend on the system aspect ratio.
This point is particularly important in the context of homogenisation. Indeed, since it describe the system as homogeneous and derives relations between averaged quantities, the boundary conditions should have as little influence as possible. This appears to be extremely difficult in this problem because the boundary layer diverges at low flow rates, but also because it depends on the aspect ratio of the system under consideration.
In the light of this study, it is difficult to claim that the boundary condition is responsible for the breakdown of the non-Newtonian behaviour observed in \cite{chevalier14}, since we still observe non-Newtonian behaviour with some fluid at rest in the boundary layer.
However, it is worth recalling that this path selection is closely related to the disorder of the system (see \cite{chevalier15a,waisbord19,kostenko19}).
It is then probably the combination of both the imposed flow rate condition and the use of a relatively homogeneous porous medium (monodisperse glass bead) that makes the detection of such phenomena difficult.

Finally, we have analysed the flow structure in the zero flow limit. We have shown that the flow structure can be mapped to the problem of an avalanche dynamics of an interface pulled at one end in a random medium.
This avalanche dynamics is characterized by power laws  whose exponent ($\tau_S$, $\tau_\ell$, $\delta$, $\zeta$) can be predicted. 
However, the analogy cannot predict the two exponents $\beta$ and $\mu$, which correspond respectively to the increase in flow rate with pressure and the increase of the boundary layer with flow rate. 

There are various possibilities for extending this study.
Firstly, this problem could be extended to a 3D pore network. In this case, the physics is certainly similar, where the inlet flow channels merge at low flow rates. However, one can expect differences in the scaling power exponents. 
One could also consider other flow driving conditions such as the application of a homogeneous body force like gravity.
If this condition should not differ too much from the pressure imposed BC, one could note that there is also a boundary layer associated with it, as can be seen on the right side of the figure \ref{fig:nchannel_depth} (right). It would be interesting to analyse the evolution of this boundary layer.

\section*{Acknowledgements}

The authors thanks F. Lanza, S. Sinha and Ph. Coussot  for fruitfull discussions. This work was partly supported by the Research Council of Norway through its Center of Excellence funding scheme, project number 262644. Further support, also from the Research Council of Norway, was provided through its INTPART program, project number 309139.

\end{document}